# The Role of Intermediate States in Low-Velocity Friction between Amorphous Surfaces

Woo Kyun Kim

*Department of Aerospace Engineering and Mechanics, the University of Minnesota, Minneapolis, Minnesota, 55455, USA*

Michael L. Falk

*Department of Materials Science and Engineering, Johns Hopkins University, Baltimore, Maryland, 21218, USA*

*Department of Mechanical Engineering, Johns Hopkins University, Baltimore, Maryland, 21218, USA*

*Department of Physics and Astronomy, Johns Hopkins University, Baltimore, Maryland, 21218, USA*

**Simulated sliding between an oxidized silicon tip and surface over six decades of velocity using accelerated molecular dynamics reproduces the experimentally observed velocity dependence of the friction force. Unlike in the crystalline case, as increasing force is applied to the amorphous tip intermediate states arise. These intermediate states serve as critical transition pathways. The emergence of such states leads to the emergence of a plateau in sliding velocity at lower sliding speeds and higher temperatures. A simple theory based on these observations successfully describes both the experimental and the simulated data.**

Friction has presented a challenge since man's earliest technological feats and has been studied on multiple length scales extending from nanometers for a protein motor [1] to several kilometers for earthquakes [2]. The atomic-scale origin of friction has also been a growing concern as nanotechnology, particularly micro/nano electro-mechanical systems (MEMS/NEMS), advances. In recent years, it has become possible to study nanoscale friction via Atomic Force Microscopy (AFM) [3]. A typical AFM tip is a single asperity contact with a radius of 10 to 100 nm [4]. Macroscopic friction arises from many such asperities, and as such AFM studies have been used to study superlubricity [5-7], the temperature and sliding velocity dependence of friction [8, 9], and the effects of surface vibrational frequency shifts on nanoscale friction [10].

Several theoretical models have been developed to interpret the experimental results [11-16], and some simulations have been performed based on these models [13, 15, 16]. However, most of these models assume idealized crystalline surfaces [11-14] although many if not most sliding surfaces are disordered due to native oxides that form spontaneously or due to the extremes of pressure and shear that accompany friction. Recently, particular attention has been brought to the fact that transitions that take place in such a system may be significantly more complex than accounted for in the canonical theories of friction [15, 17]. Interesting statistical models for considering the relative roles of bonding and de-bonding have been proposed [16], but such theories would benefit from atomic-scale analysis of the tip-surface dynamics of particular systems to confirm the mechanisms proposed are, in fact, the most critical.

Directly observing buried sliding interfaces *in situ* is still a formidable challenge. Molecular dynamics (MD) simulations [18] have been proposed as an alternative means of observing atomic behaviors [e.g., 19-22], but MD's sub-microsecond time scale limits such simulations to sliding speeds orders of magnitude faster than most AFM experiments. One exception is the extension of the parallel replica method to driven systems [23], which has reduced sliding velocities as low as 1 mm/sec. However, this method is generally insufficient to reach experimental sliding speeds in AFM studies, typically in the µm–nm/sec range. We demonstrate here that much lower sliding speeds can be achieved using an accelerated MD scheme [24] based on hyperdynamics [25, 26]. In this method, which was developed by the authors [24] and applied to a realistic problem for the first time here, potential energy function is modified in a controlled way to facilitate transitions and many such simulations are performed simultaneously in order to exploit parallel computing architectures.

Friction between an oxidized silicon AFM tip and surface under UHV conditions has been shown to exhibit unexpected relationships between friction force and sliding velocity with nontrivial temperature dependences [8]. At lower temperatures, the frictional force increases logarithmically with sliding velocity, but at higher temperatures the frictional force had no apparent dependence on the sliding velocity. The simple single-step activated transition picture, the Tomlinson model [11, 12], has been shown not to hold at low sliding speeds where multi-step transition mechanisms must be involved [17]. However, to the best of our knowledge, there has not been any direct observation of the specific mechanisms that are present during the sliding although many candidate mechanisms have been proposed [15-17]. In this study we investigate the origin of these deviations from logarithmic sliding rate dependences predicted by the Tomlinson model using our accelerated MD methodology.

First, we present the simulation results, and show that these results mirror the experimental observations [8]. Figure 1 shows our 3-dimensional AFM model consisting of an oxidized silicon tip and substrate. The spherical tip with a radius of 2.1 nm consists of 569 silicon and 309 oxygen atoms, and the substrate (4.61 nm ×4.61 nm) contains 1,152 silicon and 1,248 oxygen atoms. The atomic interactions are modeled by a modified Stillinger-Weber potential developed by Watanabe *et al*. [27]. The parameters of this potential were chosen by performing a best fit to the *ab initio* calculations of various silica clusters, and the Watanabe potential has been used to reproduce the structural features of thin oxide films [27, 28]. The oxidized layers were generated by a procedure proposed by Dalla Torre *et al*. [29], in which oxygen atoms are inserted one by one into Si-Si bonds. The tip is joined to the substrate in the [001] direction by a normal force of 1.5 nN. The atoms on the bottom of the substrate are fixed, and those on the top of the tip move like a rigid body. These atoms on the top of the tip are pulled by a spring with a stiffness of 6.1 N/m, representing an AFM cantilever. The normal force and the spring stiffness are derived from experiments in [8]. Temperature is controlled by a Nosé-Hoover chain thermostat [30], and the equations of motion are solved using a modified velocity-Verlet algorithm [31]. To reduce the sliding velocity we used a version of Voter's hyperdynamics method [25, 26] implemented specifically for simulating frictional sliding by exploiting parallel computation and transition state theory, as described in our previous work [24]. Simulations were performed at 100K and 400K spanning a wide range of sliding velocities, from 273 nm/sec to 273 mm/sec.

In Fig. 2(a) the system exhibits a non-uniform stick-slip behavior arising from the amorphous character of the surfaces, in contrast to the uniform stick-slip motion observed in MD simulation studies of crystals [23, 24]. In the simulated results, as in the experiments, we clearly observe deviation from the logarithmic velocity dependence of the friction force predicted in the Tomlinson model. Figure 2(b) shows a detailed comparison between the transitions that initiate sliding at T = 100 K and $v_s$ = 27.3 mm/sec and at T = 400 K and $v_s$ = 2.73 µm/sec. Transitions occur among three states, an initial state, A, a final state, C, and an intermediate state, B. The atomic configurations corresponding to these states are illustrated in Fig. 3(a). In state A there is one silicon-silicon bond between the tip and the substrate, and in state B the silicon atom is bonded at a different angle with the same substrate atom. In state C, this silicon atom forms a new bond with a different silicon atom in the substrate. The transition from B to C induces a larger drop in the lateral force than the transition from A to B. While at low temperature the system undergoes two one-way transitions (first from A to B and then from B to C), at high temperature the system switches back-and-forth between A and B, confirming the importance of multi-step processes as noted in previous investigations by others [15, 17]. In fact the frequency of the transitions was greater than apparent in Fig. 2(b) since limited resolution requires us to show time-averaged data.

Figs. 3(b) and 3(c) show the tip positions and the minimum energies of each state as a function of the slider position, and illustrate that the stability of each state depends on the slider position. The intermediate state B, in particular, does not emerge until a finite force is applied. This is a significant additional deviation from the Tomlinson model not anticipated in prior work such as [15] and [17]. At the smaller slider positions only states A and C are stable. As the slider moves, state B emerges and it becomes more stable than state A after a slider displacement of 0.26 nm. As the slider moves further, state A becomes unstable and states B and C remain. Neither the transition from A to C nor the backward transition from C to B is observed. The slip transition from B to C occurs at a much smaller slider position in the higher temperature case.

The observations above, taken together, imply that switching between states A and B has a significant effect on the lateral force in large part because of the intermediate nature of state B and its absence at low applied loads. If the transition occurs when only states B and C exist, as would happen at high sliding speeds and/or low temperatures, we should see the predicted logarithmic dependence [12]. At sufficiently low sliding speed and/or high temperature, A and B coexist. As the sliding speed decreases and/or the temperature rises the probability of residing in the intermediate state B decreases, and the rate of a transition to state C consequently decreases leading to a deviation from the logarithmic dependence. Below a critical force the absence of state B precludes any transitions along this pathway.

To analyze this phenomenon quantitatively we have developed a theoretical model. We begin by writing a set of Kramers' rate equations [32] amongst states A, B, and C. Using the linear relationship between the lateral force and the slider position ($F = k_{eff} x_s = k_{eff} v_s t$), we obtain the rate equations in terms of the lateral force as in Gnecco's model [12]

$$\frac{dp_A}{dF} = -\frac{R_{A \to B}(F)}{k_{eff} v_S} p_A + \frac{R_{B \to A}(F)}{k_{eff} v_S} p_B, \tag{1}$$

$$\frac{dp_B}{dF} = +\frac{R_{A \to B}(F)}{k_{eff} v_S} p_A - \left(\frac{R_{B \to A}(F) + R_{B \to C}(F)}{k_{eff} v_S}\right) p_B, \tag{2}$$

and

$$\frac{dp_C}{dF} = +\frac{R_{B \to C}(F)}{k_{eff} v_S} p_B, \tag{3}$$

where $p_A$, $p_B$, and $p_C$ are the probabilities that the system stays at A, B, and C respectively, and $R_{A \to B}$, $R_{B \to A}$, and $R_{B \to C}$ are the transition rates. We ignore the backward transitions from C to A or B and the transition from A to C. We write the rate constants in the standard Arrhenius form, $f \exp(-\beta \Delta E)$ where $\beta = 1/k_B T$. We make the approximation that $R_{A \to B}$, $R_{B \to A} \gg R_{B \to C}$ so that the system, on average, makes many more transitions between A and B before making the transition from B to C leading to the relation

$$\frac{p_B}{p_A + p_B} \approx \frac{R_{A \to B}}{R_{A \to B} + R_{B \to A}} = \frac{1}{1 + \eta(F)\exp[\beta(V_B - V_A)]} , \qquad (4)$$

where $V_A$ and $V_B$ are the energies of state A and B. The function $\eta(F)$ is the ratio of the attempt frequencies of the B→A and A→B transitions. We approximate this ratio as

$$\eta(F) = C_1 \left( \frac{F - F_U}{F - F_L} \right)^2 , \qquad (5)$$

where the A→B and B→A attempt frequencies must vanish at some lower critical force, $F_L$, and upper critical force, $F_U$, respectively since only state A is stable when $F < F_L$, only state B is stable when $F > F_U$, and both are stable when $F_L < F < F_U$. To obtain an explicit expression for the most probable force at transition, $F^*$, we assume that the transition energies vary linearly with the applied force such that $\Delta E_{B \to C} = \lambda (F_C - F)$, where $F_C$ is the lateral force when the energy barrier from B to C vanishes. We also parameterize the energy difference between states A and B in terms of the applied force by the linear expression $V_B - V_A = C_2 F + C_3$. From Eqs. (3), (4), and (5), the probability distribution function of the lateral force at transition, $g(F)$, is given by

$$g(F) = -\frac{d(p_A + p_B)}{dF} = \frac{dp_C}{dF} = \frac{R_{eq}}{k_{eff} v_S} \times (p_A + p_B) \qquad (6)$$

where

$$R_{eq} = \frac{R_{B \to C} \times R_{A \to B}}{R_{A \to B} + R_{B \to A}} = \frac{f_{B \to C} \exp[-\beta \Delta E_{B \to C}]}{1 + \eta \exp[\beta(V_B - V_A)]} . \qquad (7)$$

Then, the most probable lateral force at transition, $F^*$, satisfying $g'(F^*) = 0$ is obtained by solving the following equation

$$\left. \frac{dR_{eq}}{dF} \right|_{F^*} = \frac{(R_{eq})^2}{k_{eff} v_S} . \qquad (8)$$

By imposing this condition and incorporating the relationships $\Delta E_{B \to C} = \lambda (F_C - F)$ and $V_B - V_A = C_2 F + C_3$, we can derive the sliding speed as a function of the temperature and friction force,

$$\frac{v_S}{v_O} = \frac{\exp[\lambda \beta (F^* - F_C)]}{1 + \left[\left(1 - \frac{C_2}{\lambda}\right)\eta - \frac{\eta'}{\lambda \beta}\right]\exp[\beta(C_2 F^* + C_3)]}, \quad (9)$$

where

$$v_O = \frac{f_{B \to C}}{\lambda \beta k_{eff}}. \quad (10)$$

Note that when only states B and C exist, we recover a logarithmic dependence as in Gnecco's model [12] and Eq. (9) simply becomes $F^* = (1/\lambda\beta) \ln (v_s / v_o) + F_C$.

Now, we analyze both the simulation results and the experimental observations using the master Eq. (9). The comparison of the theory and the simulation is shown in Fig. 4(a) using fit parameters from Table I. Most parameters are determined directly from Fig. 3 and the high-velocity portion of the data leaving only one independent fitting parameter. It is apparent that the switching between states A and B, modeled in the denominator of Eq. (9), results in a plateau at the higher temperature. We expect this theory is too simple to accurately model the experiments that represent an average over many dissimilar transitions. However, if such state switching is common, and if the energies associated with such transitions are not broadly distributed, the model should be able to capture the general trends in the data. Figure 4(a) also shows the force at the initiation of sliding in comparison with the experimental data from [8]. The forces measured in simulation are the peak force during a single slip transition while the experiments measure average sliding force over several nanometers. For this reason we expect the two quantities to be related by a factor of 2.

Figure 4(b) shows an application of the theory to describe the lateral forces measured by AFM in [8] over a wide range of temperatures (55 K ~ 255 K) and at the sliding velocities ranging from 100 nm/sec to 16 µm/sec. The fit parameters used in Fig. 4(b) are given in Table I. The general trend of decreasing force and decreasing velocity dependence with increasing temperature is captured by the theory. More importantly, the transition from logarithmic dependence at low temperature to a high temperature plateau is consistent with predictions based on the theory. The non-monotonic dependence of force on temperature at high temperature we believe may arise from experimental limits accurately resolving such small differences in force. The model also predicts that at higher sliding velocity the logarithmic dependence will be recovered at higher temperatures.

In conclusion, we have shown using a novel application of accelerated MD that switching between states, combined with the emergence of new intermediate states during sliding, together play a crucial role in determining the forces during sliding contact between amorphous surfaces. This work demonstrates the impact that simulation methodologies capable of reaching longer time scales can have in the theoretical interpretation of experimental investigations on the nanoscale. Moreover, it is striking that even a rather simple theoretical model that includes the observed mechanism is able to explain the deviation from the logarithmic dependence analytically over a wide range of sliding speeds and temperatures in an actual experimental system. We conclude that this type of transition must be quite common in frictional sliding when one or more of the involved surfaces are not perfectly crystalline and, in turn, our model is more applicable than the Tomlinson-derived models of friction typically assuming the uni-directional transition mechanism.

We acknowledge support of the NSF program on Materials and Surface Engineering under Grants CMMI-0510163 and CMMI-0926111 and the use of facilities at the Johns Hopkins University Homewood High Performance Compute Cluster. We also thank A. Schirmeisen and M.O. Robbins for useful discussions.

TABLE I. Fitting parameters used in Figs. 4(a) and 4(b).

| Parameters | Simulation (Fig. 4(a)) | Experiment (Fig. 4(b)) |
|---|---|---|
| $\lambda$ (Å) | 0.828 | 0.302 |
| $F_C$ (nN) | 1.09 | 1.15 |
| $F_L$ (nN) | 0.25 | 0.1 |
| $F_U$ (nN) | 0.6 | 0.6 |
| $F_C f_{B \to C} / k_{eff}$ (m/sec) | 142 | 572 |
| $C_1$ | 120 | 10 |
| $C_2$ (eV/nN) | -1.80 | -0.25 |
| $C_3$ (eV) | 0.681 | 0.01 |

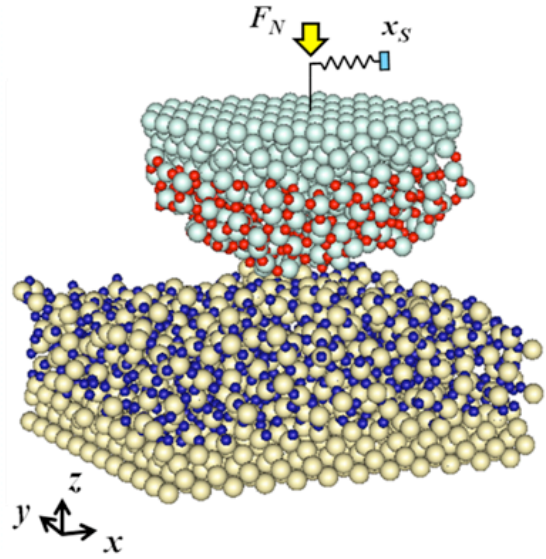

FIG. 1. A diagram of 3-dimensional AFM model consisting of an oxidized silicon tip and a substrate. Dark-colored smaller spheres are oxygen atoms and light-colored larger spheres are silicon atoms.

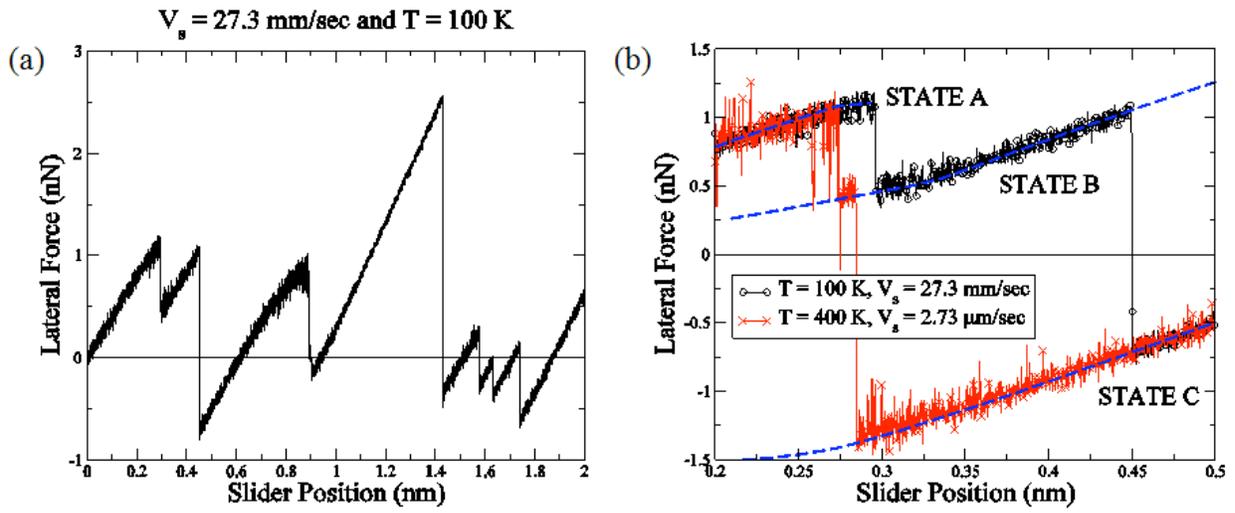

FIG. 2. (a) The lateral force as a function of slider position measured at a sliding velocity of 27.3 mm/sec and a temperature of 100 K. (b) Lateral forces near the first transition as functions of the slider position. Open circles show the data at T = 100 K and Vs = 27.3 mm/sec and crosses show the data at T = 400 K and Vs = 2.73 μm/sec. Dashed lines represent the three states illustrated in Fig. 3.

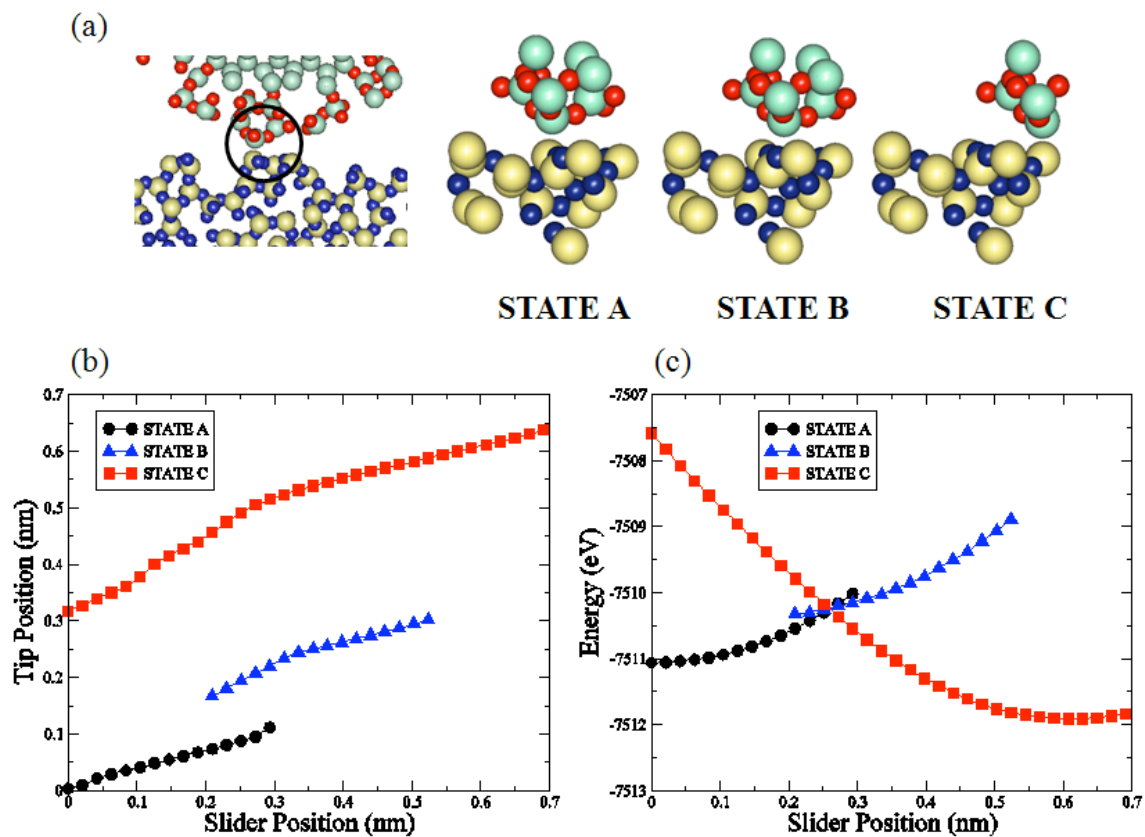

FIG. 3. Atomic configurations of state A, B, and C and the tip positions and energies at each state. (a) In state A and B, the same silicon atoms in the tip and the substrate are bonded at different bond angles. In state C, the silicon atom in the tip is bonded with another silicon atom in the substrate silicon atom. (b) The tip positions of the minimized configurations corresponding to each state as functions of the slider position. (c) The minimum energies of each state as functions of the slider position.

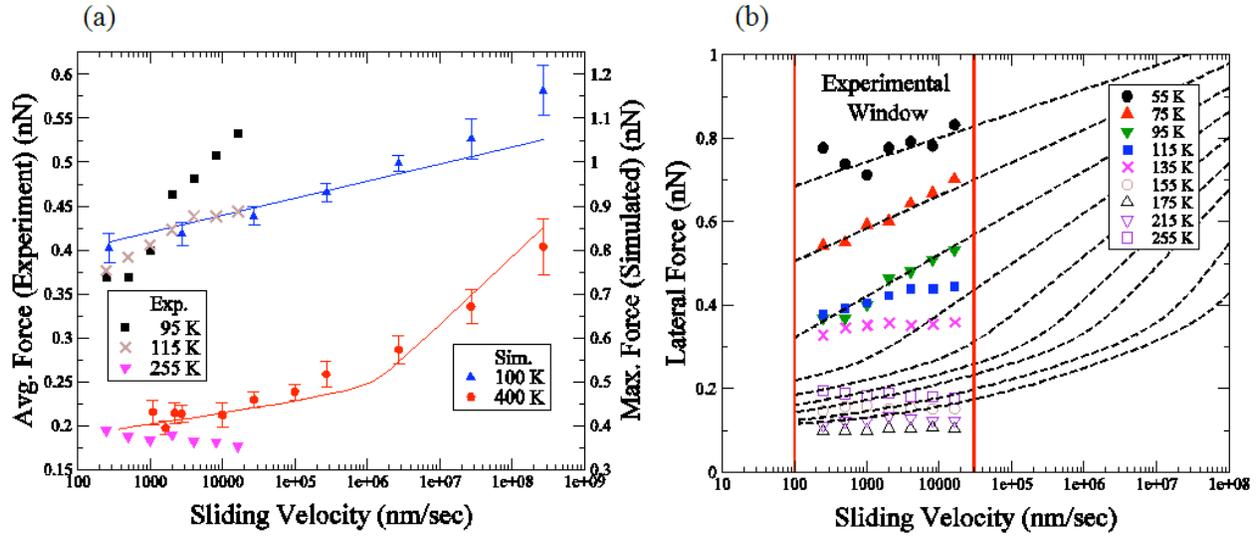

FIG. 4. Lateral forces as functions of sliding velocity from both experiments (Ref. [8]) and simulations with fitting curves constructed from the intermediate state switching model. (a) The lateral forces in the experiments (95K, 115K ,and 255 K) are average forces and the lateral forces in the simulations (100K and 400 K) are measured at the transition from state B to state C. At 100 K the logarithmic dependence of the lateral force extends up to low velocity regime, but at 400 K the plateau region is observed below a sliding velocity of 10 μm/sec. The error bars represent the standard deviations. (b) Lateral forces retrieved from an AFM experiment over a wide range of temperatures and at sliding velocities are shown in the experimental window. Solid curves in (a) and dashed curves in (b) are fits based on the intermediate state switching theory discussed in the text.